\documentclass{aa}  

\usepackage{graphicx}
\usepackage{txfonts}
\usepackage{lipsum}
\usepackage{subcaption}
\usepackage{lscape}
\usepackage{placeins}
\usepackage{rotating}
\usepackage{soul}
\usepackage{float}

\begin{document}

   \title{Statistical structure and physical interpretation of the fillout factor distribution in contact binary stars}
   \titlerunning{Fillout factor distribution in contact binary stars}

   \author{A. Poro\inst{1,2}
        \and K. Li\inst{3}
        \and E. Paki\inst{4}
        \and F. Alicavus\inst{5,6}
        \and N. Alan\inst{7}
        }

   \institute{LUX, Observatoire de Paris, CNRS, PSL, 61 Avenue de l'Observatoire, 75014 Paris, France\\
             \email{atila.poro@obspm.fr}
            \and Astronomy Department, Raderon AI Lab., BC., Burnaby V5C 0J3, Canada
            \and Shandong Key Laboratory of Space Environment and Exploration Technology, Institute of Space Sciences, School of Space Science and Technology, Shandong University, Shandong 264209, People's Republic of China
            \and Binary Systems of South and North (BSN) Project; Independent Researcher, 15875 Tehran, Iran
            \and Çanakkale Onsekiz Mart University, Faculty of Science, Department of Physics, 17020, Çanakkale, Türkiye
            \and Çanakkale Onsekiz Mart University, Astrophysics Research Center and Ulupnar Observatory, 17020, Çanakkale, Türkiye
            \and Fatih Sultan Mehmet Vakif University, Department of History of Science, 34664 Istanbul, Türkiye\\}
   \date{Received ---, ---}

   \abstract
{Contact binary systems exhibit complex evolutionary behavior governed by mass transfer, angular momentum loss (AML), and Roche geometry. The fillout factor ($f$) is widely used as a descriptor of the degree of overcontact, yet its physical and statistical role remains nonuniformly defined across the literature. We used a compiled observational sample of W UMa-type contact binaries to construct a data-driven classification framework based on $f$ and its relation to fundamental system parameters. Unsupervised clustering applied to the one-dimensional $f$ distribution reveals a robust three-class structure corresponding to shallow-, medium-, and deep-contact configurations. The statistically determined boundaries are located at $f \simeq 0.257$ and $f \simeq 0.561$, with a stable silhouette score of $S \simeq 0.613$. The stability of this classification is confirmed through perturbation tests in $f$ and parameter-conditioned samples. A random forest regression model shows that 60.7\% of the variance in $f$ can be explained by the considered physical and geometric system parameters, with geometric parameters providing the largest predictive contribution. Among the considered parameters, the mass ratio emerges as the dominant predictor. The identified subclasses therefore appear to reflect a continuous geometric sequence rather than discrete physical states, governed by the progressive evolution of the Roche equipotential configuration under mass transfer and AML. Notably, the second boundary separating medium- and deep-contact systems shows lower stability than the first boundary, suggesting that the transition toward deep contact may be more sensitive to the physical composition of the analyzed sample.}

   \keywords{binaries: close -- stars: evolution -- methods: statistical, data analysis
               }

   \maketitle
   \nolinenumbers

\section{Introduction}
Contact binary systems represent one of the most fascinating and physically complex configurations in stellar astrophysics. These systems generally consist of two late-type stars that share a common convective envelope (CCE), filling their Roche lobes and orbiting each other with periods ($P$) typically shorter than one day (\citealt{1941ApJ....93..133K}, \citealt{1968ApJ...151.1123L}). The fillout factor is a key geometric parameter that quantifies the extent of physical contact between the two stellar components. It is defined as the degree to which the stellar surfaces exceed the inner critical Roche lobe, expressed by the following equation:
\begin{equation} 
f = {\Omega-\Omega_{in}\over \Omega_{out}-\Omega_{in}},
\end{equation}
where $\Omega$ is the surface potential of the binary, while $\Omega_{in}$ and $\Omega_{out}$ denote the inner and outer Roche lobe potentials, respectively. By construction, 
$f=0$ corresponds to a system in which both components exactly fill their Roche lobes (marginal contact), while increasing positive values of $f$ indicate progressively deeper immersion of the stellar surfaces into a common envelope. In practice, the fillout factor is usually derived by light curve modeling using codes such as the Wilson‑Devinney program (\citealt{1971ApJ...166..605W}, \citealt{1979ApJ...234.1054W}), PHysics Of Eclipsing BinariEs (PHOEBE; \citealt{2005ApJ...628..426P}, \citealt{2016ApJS..227...29P}), and more recently, the BSN application (\citealt{2025Galax..13...74P}), where it is treated as a free parameter coupled with other physical quantities including the mass ratio ($q$), orbital inclination ($i$), component temperatures ($T_{1,2}$), and fractional radii ($r_{1,2}$).

Despite its wide usage, the fillout factor suffers from nonuniform classification criteria in the literature. While most authors qualitatively distinguish between shallow-, medium-, and deep-contact configurations, the numerical thresholds adopted vary considerably. An empirical three-category scheme has emerged: shallow contact ($f < 0.25$), medium contact ($0.25 < f < 0.5$), and deep contact ($f > 0.5$) (e.g., \citealt{2022AJ....164..202L}, \citealt{2023MNRAS.519.5760L}). However, this scheme is purely empirical and lacks any theoretical foundation. Different authors propose different thresholds according to their specific samples or scientific goals. For instance, \cite{2020RAA....20..163Q} provided a comprehensive review of contact binary populations observed by the Large Sky Area Multi-Object Fiber Spectroscopic Telescope (LAMOST; \citealt{2012RAA....12.1197C}, \citealt{2015RAA....15.1095L}), and discussed the evolutionary stages of these systems from a physical perspective, and explicitly defined marginal (shallow) contact as $f<20\%$ and deep contact as those with $f\geqslant50\%$, leaving the intermediate range between 20\% and 50\% as an unlabeled transition zone. Although the qualitative distinction is conceptually consistent across studies, the precise thresholds differ by up to 5–10 percentage points, creating difficulties when comparing evolutionary studies or interpreting the physical significance of a given $f$ value. This difficulty is particularly acute for systems such as deep, low-mass-ratio contact binaries (DLMCBs, e.g., GR Vir with $q=0.122$ and $f=78.6\%$; \citealt{2004AJ....128.2430Q}), which represent the final evolutionary stage before a merger. The existence of such physically meaningful subclasses underscores the need for a statistically robust classification framework that can objectively identify them without relying on arbitrarily chosen thresholds.

From a physical perspective, the fillout factor is expected to be related to fundamental system properties. Theoretical models of contact binary evolution (\citealt{1982A&A...109...17V}, \citealt{1995MNRAS.274.1019S}, \citealt{2005ApJ...629.1055Y}) suggest that AML, together with mass transfer, can lead to the gradual expansion of the common envelope, resulting in a secular increase in the fillout factor. Meanwhile, thermal relaxation oscillation (TRO) models (\citealt{1994ApJ...434..277W}, \citealt{2001MNRAS.328..914Q}, \citealt{2025ApJ...995...19F}) propose cyclic variations in the degree of contact due to alternating phases of thermal disequilibrium. However, large‑scale statistical tests of these predictions have remained limited, partly because a robust, data‑driven classification of $f$ has been lacking.

Beyond its passive geometric description, the fillout factor is expected to be coupled to the long-term evolution of contact binaries through both angular momentum loss (AML) and energy transfer (ET) processes. As shown by the TRO models of \cite{1976ApJ...205..217F}, \cite{1976ApJ...205..208L}, \cite{2004MNRAS.355.1383L}, \cite{2005ApJ...629.1055Y}, and \cite{2025ApJ...995...19F}, the cyclic exchange of energy and mass between the components can naturally lead to oscillations in the degree of contact. Furthermore, a strong relation between $f$, $q$, and $P$ arises naturally from the Roche geometry and Kepler's third law: for a system with a given primary radius and mass ratio, a shorter orbital period necessarily requires a larger fillout factor. Since AML drives a secular decrease in the orbital period, it naturally produces a long-term increase in the fillout factor. In contrast, ET reverses the direction of mass transfer and, in most models, tends to produce a secular decrease in the mass ratio that, at a constant total mass, increases the orbital period and partially counteracts the AML-driven trend. Consequently, the long-term evolution of $f$ reflects the competing influences of AML and ET on the orbital evolution of the system. Moreover, because the TRO model predicts cyclic variations in the fillout factor, a given value of $f$ cannot always be uniquely associated with a specific evolutionary stage.

The recognition of $f$ as both a geometric and evolutionary parameter underscores the need for a statistically robust, observationally driven classification free from arbitrary thresholds. In this context, we treat the fillout factor as an intrinsic evolutionary coordinate and investigate whether its observed distribution alone contains statistically meaningful structure that can be linked to underlying physical parameters. Accordingly, the present study has three main objectives. First, we establish a statistically motivated, observer-independent classification of the fillout factor using a larger sample of contact binaries, without imposing predefined arbitrary thresholds. Second, we quantify the relative predictive contributions of geometric parameters and thermal parameters to the observed distribution of $f$. Third, we examine whether the statistically derived classes correspond to distinct physical states or rather represent segments of a continuous evolutionary sequence.

\section{Dataset}
A statistically robust and data-driven classification scheme was developed using a sample of 796 contact systems from \cite{2025MNRAS.538.1427P}. The sample was restricted to systems for which the fillout factor and all other parameters required for the present analysis were available in the catalog. Consequently, the final sample size is smaller than the original sample reported in \cite{2025MNRAS.538.1427P}. The adopted dataset provides the fillout factor, orbital period, mass ratio, component temperatures, and fractional radii of both stars. Mass ratios reported in the literature with $q>1$ were converted to their reciprocal values prior to the statistical analysis.

The final dataset comprises 51.1\% A-subtype and 48.9\% W-subtype contact binary systems. In terms of eclipse morphology, 54.3\% of the systems exhibit partial eclipses, whereas 45.7\% are total-eclipsing binaries. Most published solutions (80.3\%) are based solely on photometric light-curve analyses, while the remaining 19.7\% include spectroscopic constraints. Third light was not required in 86.7\% of the published models, whereas a nonzero third-light contribution was reported for 13.3\% of the systems. Likewise, 51.4\% of the published solutions include one or more starspots, while 48.6\% were modeled without spots.

The parameter distributions span a broad range of contact binary properties. The orbital periods extend from 0.1793 to 1.1494 d, mass ratios from 0.036 to 1.000, and fillout factors from 0.00 to 0.98. The adopted primary and secondary temperatures range from 3600 to 9600 K and from 3135 to 9184 K, respectively. The fractional radii cover intervals of $0.259 \le r_1 \le 0.857$ and $0.156 \le r_2 \le 0.546$. Density-strip representations of the parameter distributions together with the pairwise parameter correlations are presented in Figures~\ref{Fig:density} and \ref{Fig:scatter}.

Since the adopted dataset was compiled from previously published studies, the statistical results presented here should be interpreted as representative of the currently available sample of analyzed contact binary systems rather than the intrinsic population of all contact binaries. Furthermore, the published solutions were derived from different observational datasets (ground-based, space-based, or a combination of both), modeling strategies, and light curve analysis codes, which may introduce a degree of heterogeneity into the compiled parameters. The adopted parameters are also not uniformly distributed over their full ranges, with some regions of the parameter space being more densely populated than others (Figure~\ref{Fig:density}).

\section{Statistical classification}
\begin{figure*}
\sidecaption
\includegraphics[width=12cm]{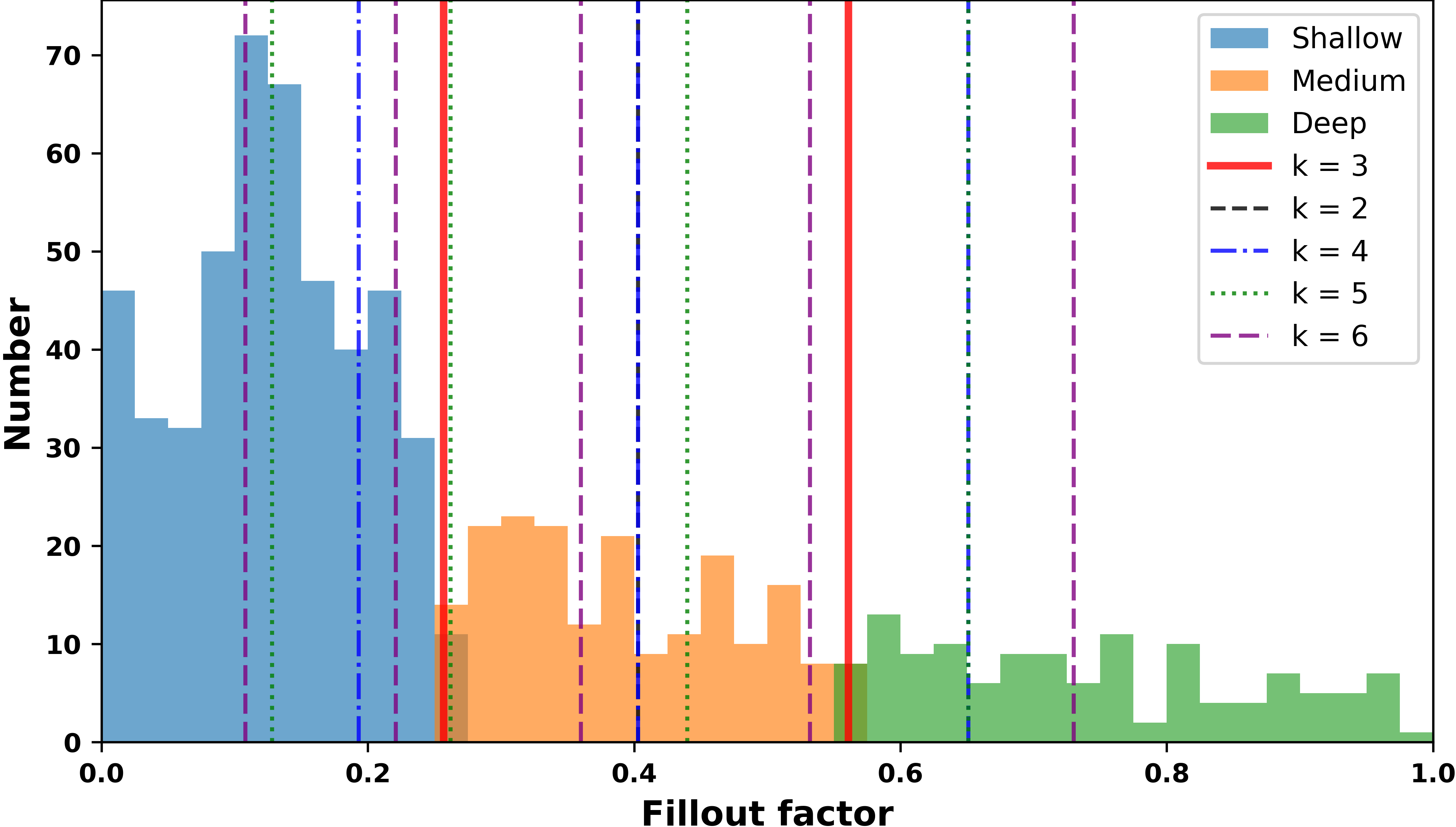}
\caption{Distribution of the fillout factor across the statistically determined subclasses of contact binaries. The three colored histograms correspond to the shallow-, medium-, and deep-contact groups obtained from the data-driven classification with $k=3$. The vertical lines indicate the cluster boundaries obtained from different choices of the number of clusters ($k=2-6$).}
\label{Fig:f_classification}
\end{figure*}

The K-means clustering algorithm was applied to the one-dimensional distribution of the fillout factor, and solutions ranging from two to six clusters were evaluated using the silhouette score. This range was selected to explore whether the commonly adopted empirical classifications correspond to statistically distinguishable structures in the observed distribution while avoiding excessively fragmented partitions that become difficult to interpret physically. The silhouette score ranges from $-1$ to $+1$, with larger positive values indicating more compact and better-separated clusters, values near zero indicating overlapping clusters, and negative values suggesting that many objects are assigned to inappropriate clusters. Unlike other system parameters, which describe dynamical and thermal properties of the system, the fillout factor provides a dimensionless measure of the degree of contact, enabling direct comparisons among systems with different masses and orbital periods. This property makes the fillout factor a natural coordinate for identifying statistically meaningful structure in the observed distribution. The best partition was obtained for three clusters, which were then sorted according to their mean $f$ values to ensure physically meaningful progression from low to high contact degree. The corresponding silhouette scores for the tested numbers of clusters were $S=0.579$, $0.613$, $0.578$, $0.545$, and $0.531$ for $k=2$, 3, 4, 5, and 6, respectively, indicating a marginally superior clustering performance for the three-cluster solution. The resulting cluster boundaries for each tested value of $k$ are illustrated in Figure~\ref{Fig:f_classification}. Although the three-cluster solution yielded the highest silhouette score within the explored range, the resulting classes should be interpreted as statistically useful partitions of a continuous distribution rather than evidence for intrinsically discrete physical populations.

\begin{figure*}[ht!]
\resizebox{\hsize}{!}{\includegraphics{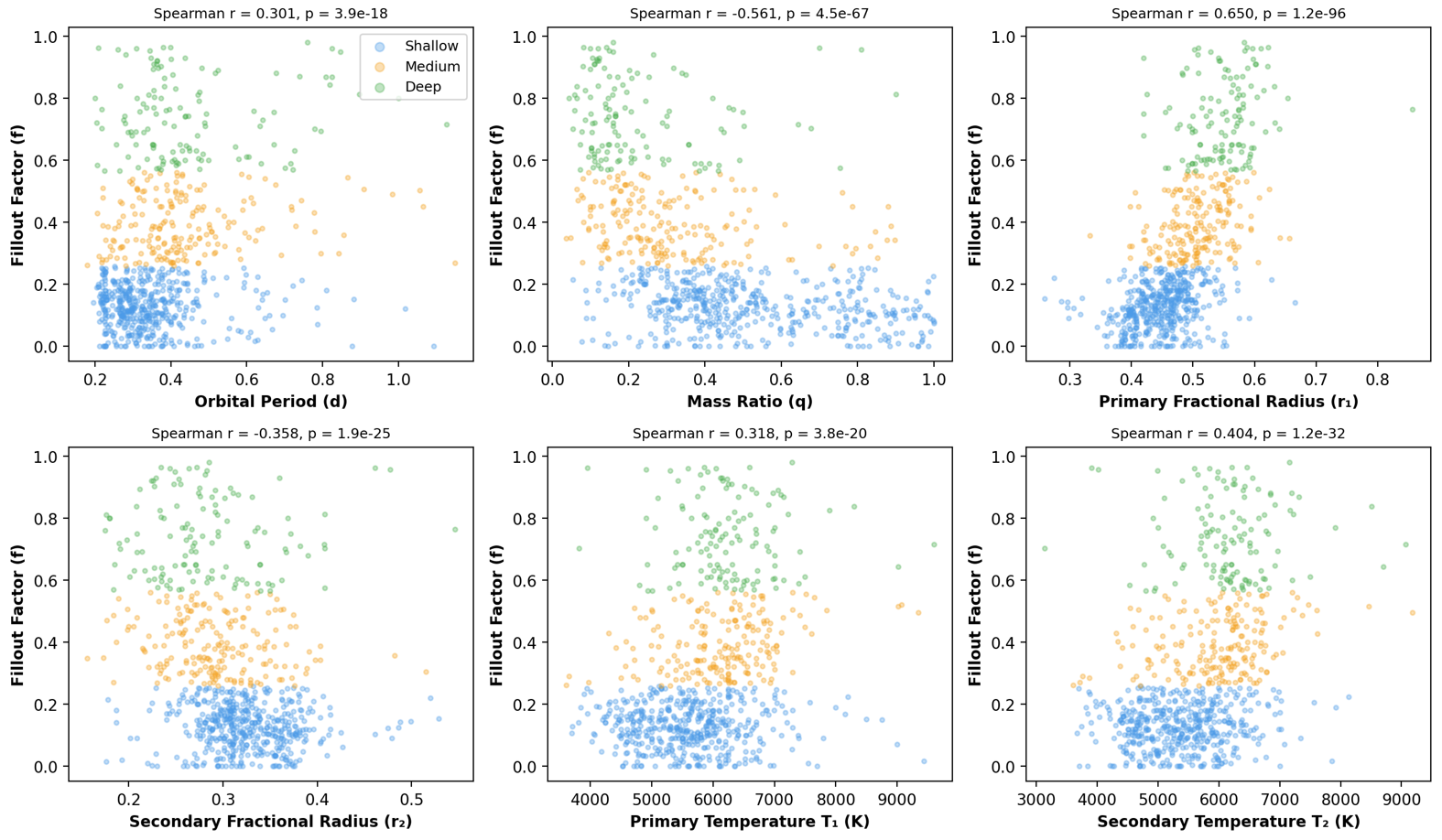}}
\caption{Fillout factor versus each physical parameter, color-coded by the derived class (blue: shallow, orange: medium, green: deep). Dashed horizontal lines indicate class boundaries at $f=0.257$ and $f=0.561$. Spearman rank correlation coefficients and $p$-values are shown for each panel, indicating statistically significant monotonic relationships between $f$ and all physical parameters.}
\label{Fig:f(b)}
\end{figure*}

The K-means algorithm was applied to the univariate distribution of $f$ rather than the full multidimensional parameter space. This choice is justified by the primary objective of the classification, which is to identify statistically meaningful and physically interpretable boundaries along the fillout factor axis itself. For comparison, K-means with the statistically preferred three-cluster solution ($k=3$) was also applied to the full six-dimensional parameter space ($P$, $q$, $r_1$, $r_2$, $T_1$, $T_2$). The resulting silhouette score evaluated in the six-dimensional space is 0.27, while the derived cluster labels yield a silhouette score of $-0.02$ when projected onto the $f$ distribution alone, indicating that clustering in the full parameter space fails to produce coherent separation along the fillout factor axis. In contrast, the univariate K-means applied to $f$ yields a silhouette score of 0.613, confirming well-separated statistical groups along the fillout factor distribution. The existence of significant monotonic relationships between $f$ and each physical parameter provides additional support for the one-dimensional approach: Spearman rank correlations were adopted because several parameters exhibit non-Gaussian distributions and potentially nonlinear monotonic trends with the fillout factor. Spearman rank correlations between $f$ and the physical parameters are statistically significant for all six quantities (Figure~\ref{Fig:f(b)}), with the largest monotonic associations found for the primary fractional radius ($r_s = 0.650$, $p = 1.2 \times 10^{-96}$), mass ratio ($r_s = -0.561$, $p = 4.5 \times 10^{-67}$), and secondary temperature ($r_s = 0.404$, $p = 1.2 \times 10^{-32}$). The strong correlation with the primary fractional radius is expected, since the fractional radii are geometrically linked to the fillout factor through Roche geometry and are therefore not fully independent quantities. These correlations indicate that the fillout factor is associated with changes in several physical properties, supporting the use of the $f$ axis as a statistically meaningful classification coordinate.

\begin{figure}[ht!]
\centering
\includegraphics[width=0.48\textwidth]{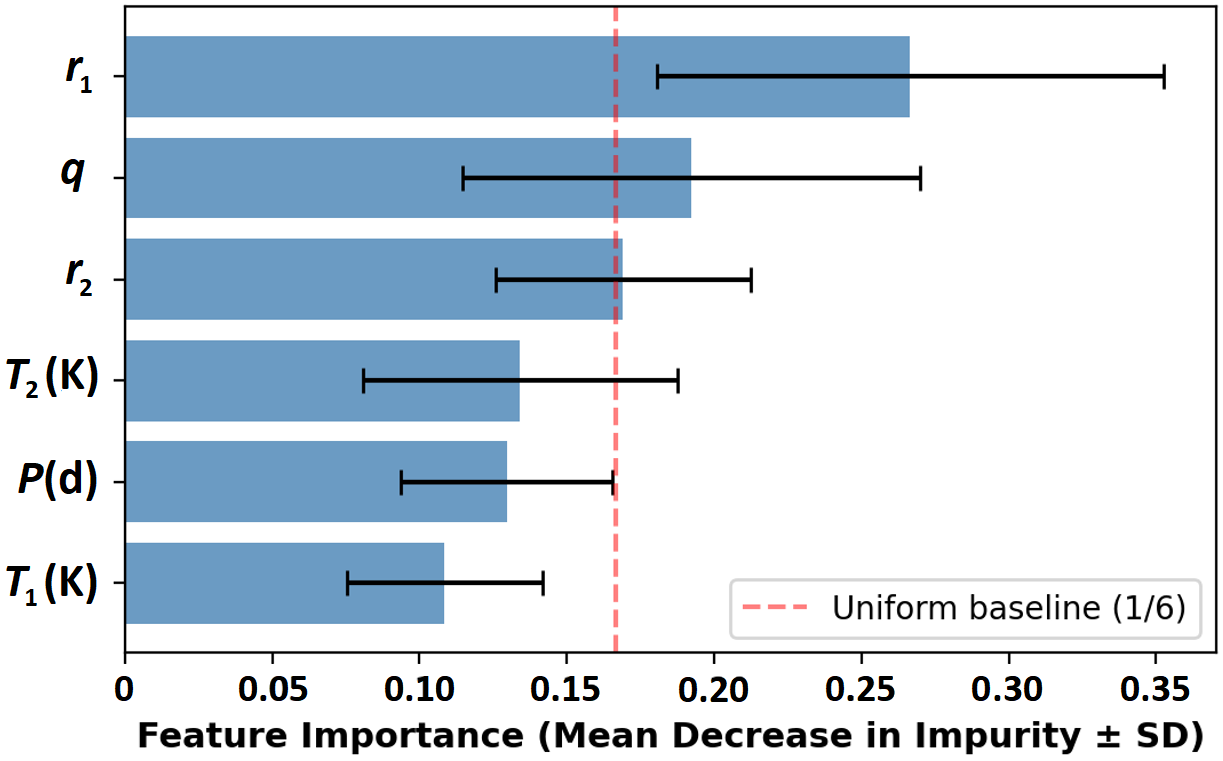}
\caption{RF feature importance values with standard deviations computed across all 500 trees in the ensemble. The dashed red line marks the uniform baseline. Parameters exceeding the baseline contribute above chance to the classification of fillout factor groups.}
\label{Fig:f(c)}
\end{figure}

\begin{figure}[ht!]
\centering
\includegraphics[width=0.48\textwidth]{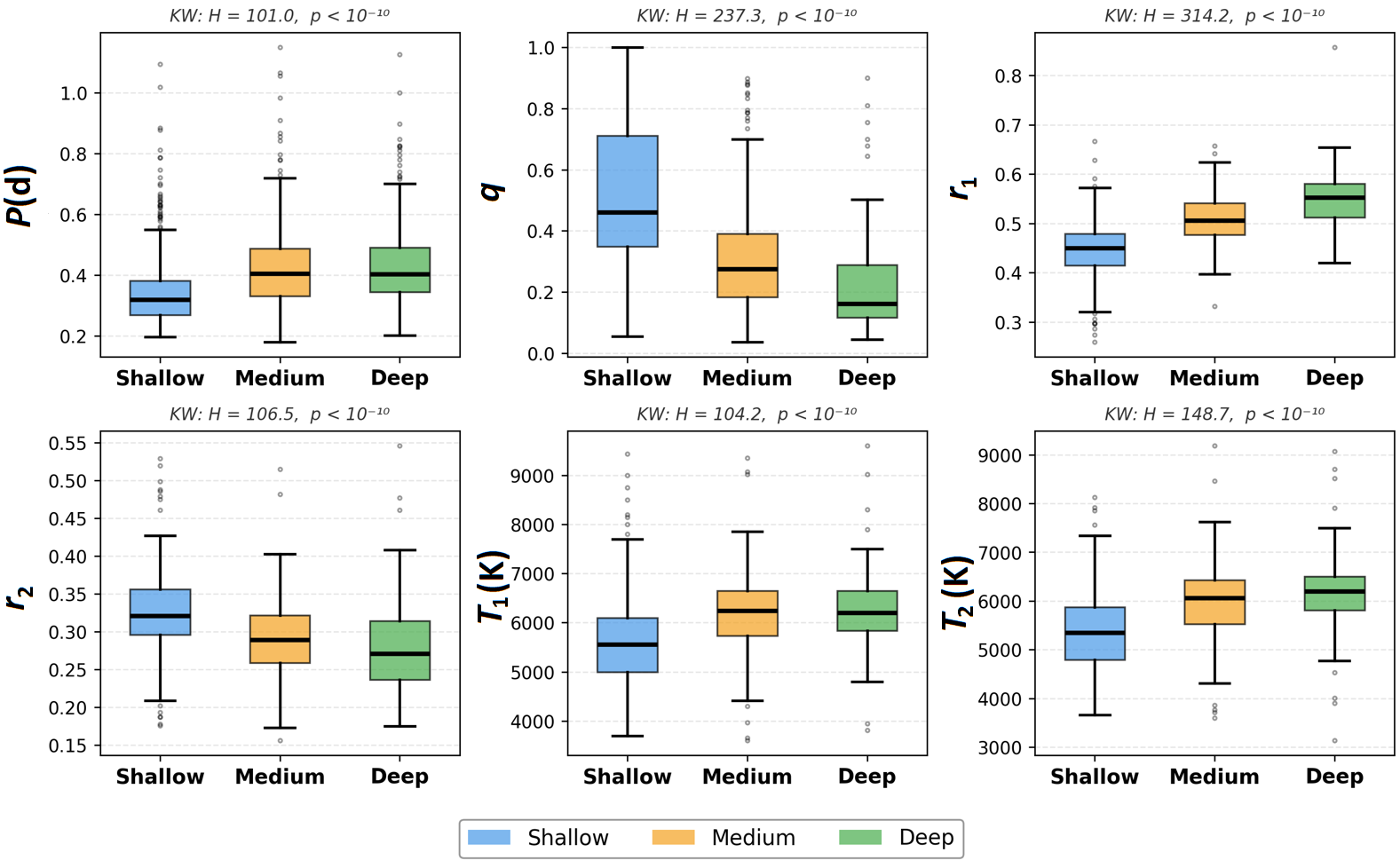}
\caption{Distribution of physical parameters across the three fillout factor classes. Each box represents the interquartile range; whiskers extend to $1.5\times$IQR. Kruskal-Wallis $H$-statistics and $p$-values above each panel confirm statistically significant differences across all parameters.}
\label{Fig:f(a)}
\end{figure}

After the fillout factor classes had been established by the K-means analysis, a random forest (RF; \citealt{2001MachL..45....5B}) classifier was used to evaluate how strongly the nongeometrical physical parameters predict membership in these statistically derived classes. To evaluate and validate the contribution of each physical parameter to these data-driven classes, we trained an RF classifier using the orbital period, mass ratio, and component temperatures. The fractional radii were intentionally excluded from this analysis because they are directly linked to the Roche geometry and are therefore not fully independent of the fillout factor itself. The inclusion or exclusion of $r_1$ and $r_2$ thus depends on the objective of each analysis: they are excluded from the classification task to avoid geometric redundancy with $f$, while being retained in regression analyses to assess the overall extent to which the fillout factor can be reconstructed from the full set of physical parameters. Prior to the RF analysis, the physical parameters were standardized to place all predictors on comparable scales. The K-means classification defines three nonoverlapping ranges of the fillout factor: systems with $0 \leq f \leq 0.257$ correspond to the shallow-contact class, those with $0.257 < f \leq 0.561$ to the medium-contact class, and those with $f > 0.561$ to the deep-contact class. The RF analysis then quantifies how well the remaining physical parameters predict membership in these predefined classes. According to the RF analysis performed using only the indirect geometrical parameters ($P$, $q$, $T_{1}$, and $T_{2}$), the mass ratio emerges as the most important predictor of class membership, contributing approximately 34.1\% of the total feature importance. The orbital period and secondary temperature provide comparable contributions (23.1\% and 22.6\%, respectively), followed by the primary temperature (20.2\%). This result indicates that the statistically derived fillout factor classes are most strongly associated with the mass-ratio distribution of the systems, while temperature-related parameters and orbital period provide secondary predictive information. It should be noted that the RF feature importance measures the predictive contribution of each parameter to the classification task and does not represent pairwise statistical correlations. Figure \ref{Fig:f(c)} shows the feature importance values with standard deviations computed across all trees in the ensemble. To verify that the three derived classes correspond to statistically distinct regions of the parameter space and not merely arbitrary numerical partitions of the fillout factor, we conducted Kruskal-Wallis nonparametric tests for each physical parameter across the three classes. The Kruskal-Wallis test was adopted as a non-parametric method for comparing independent groups because several parameters exhibit abnormal distributions and unequal variances between the derived classes. The corresponding $p$-values were computed using the standard asymptotic $\chi^2$ approximation of the Kruskal-Wallis statistic with $g-1=2$ degrees of freedom ($g=3$ classes). The results confirm statistically significant differences in all six parameters: orbital period ($H = 101.0$, $p = 1.2 \times 10^{-22}$), mass ratio ($H = 237.3$, $p = 3.0 \times 10^{-52}$), primary fractional radius ($H = 314.2$, $p = 6.0 \times 10^{-69}$), secondary fractional radius ($H = 106.5$, $p = 7.5 \times 10^{-24}$), primary temperature ($H = 104.2$, $p = 2.4 \times 10^{-23}$), and secondary temperature ($H = 148.7$, $p = 5.1 \times 10^{-33}$). The classification captures statistically meaningful trends in the parameter space, although the separation between the medium- and deep-contact groups remains weaker for several parameters. The mean values and standard deviations of all parameters for each class are summarized in Table \ref{Tab:fclass}. The effect sizes between the shallow and deep classes, quantified by Cohen’s $d$, are particularly large for the primary fractional radius ($d = -1.93$) and mass ratio ($d = 1.48$), confirming that these two parameters provide the strongest physical discrimination between the contact extremes. Pairwise post hoc Mann-Whitney tests with Bonferroni correction further reveal that the shallow class is robustly separated from both the medium and deep classes across all six parameters. The boundary between the medium and deep classes is less pronounced, with significant differences retained only for the primary fractional radius and mass ratio after correction; the remaining parameters (orbital period, secondary fractional radius, primary temperature and secondary temperature) do not show statistically significant differences between the medium and deep classes. Figure \ref{Fig:f(a)} shows the distribution of each physical parameter across the three classes with Kruskal-Wallis statistics for each panel, and Figure \ref{Fig:f(b)} shows the relationship between the fillout factor and each physical parameter with class boundaries overlaid, illustrating the overall trends of the derived groups in the observational parameter space.

Figure \ref{Fig:f_classification} illustrates the distribution of the fillout factor within the three identified categories. Each colored histogram represents one of the contact subclasses, showing the statistically derived separation between the shallow-, medium-, and deep-contact ranges in the fillout factor distribution. This approach provides a statistically motivated and internally consistent framework for future studies of contact binary evolution.

\begin{table}
\centering
\caption{Mean and standard deviation of physical parameters for the three fillout factor classes.}
\resizebox{\hsize}{!}{%
\begin{tabular}{lcccc}
\hline
Parameter & Shallow & Medium & Deep \\
& ($n=475$) & ($n=195$) & ($n=126$) \\
\hline
$f$ & $0.126\pm0.069$ & $0.391\pm0.087$ & $0.738\pm0.124$ \\
$P$ (d) & $0.344\pm0.121$ & $0.436\pm0.166$ & $0.450\pm0.177$ \\
$q$ & $0.518\pm0.230$ & $0.313\pm0.188$ & $0.223\pm0.164$ \\
$r_1$ & $0.447\pm0.050$ & $0.507\pm0.049$ & $0.549\pm0.055$ \\
$r_2$ & $0.324\pm0.047$ & $0.291\pm0.054$ & $0.282\pm0.064$ \\
$T_1$ (K) & $5593\pm850$ & $6139\pm842$ & $6230\pm803$ \\
$T_2$ (K) & $5365\pm744$ & $5969\pm796$ & $6152\pm814$ \\
\hline
\end{tabular}%
}
\tablefoot{Class sizes are given in parentheses.}
\label{Tab:fclass}
\end{table}

\section{Results and discussion}
\subsection{Sensitivity analysis of the statistical boundaries}
To evaluate the stability of the statistically derived fillout factor subclasses against small perturbations in the input distribution, we performed an additional sensitivity analysis using controlled artificial perturbations applied directly to the fillout factor values. This experiment tests whether the derived boundaries remain stable under increasing perturbations of the fillout factor distribution.

For each perturbation level, random Gaussian noise was independently added to the fillout factor of every system such that the perturbed values follow $f' = f + \delta$, where $\delta$ is drawn from a normal distribution centered at zero with a standard deviation corresponding to a selected fraction of the original fillout factor value. Perturbation amplitudes ranging from 1\% to 30\% were explored. After perturbation, all values were restricted to the physically allowed interval $0 \leq f \leq 1$. For each perturbation level, the K-means clustering procedure with three subclasses was repeated 500 times using the same methodology adopted for the original classification analysis. The statistical boundaries separating the shallow-, medium-, and deep-contact subclasses were recomputed during each realization, together with the corresponding silhouette score.

\begin{figure*}
\centering
\includegraphics[width=0.495\textwidth]{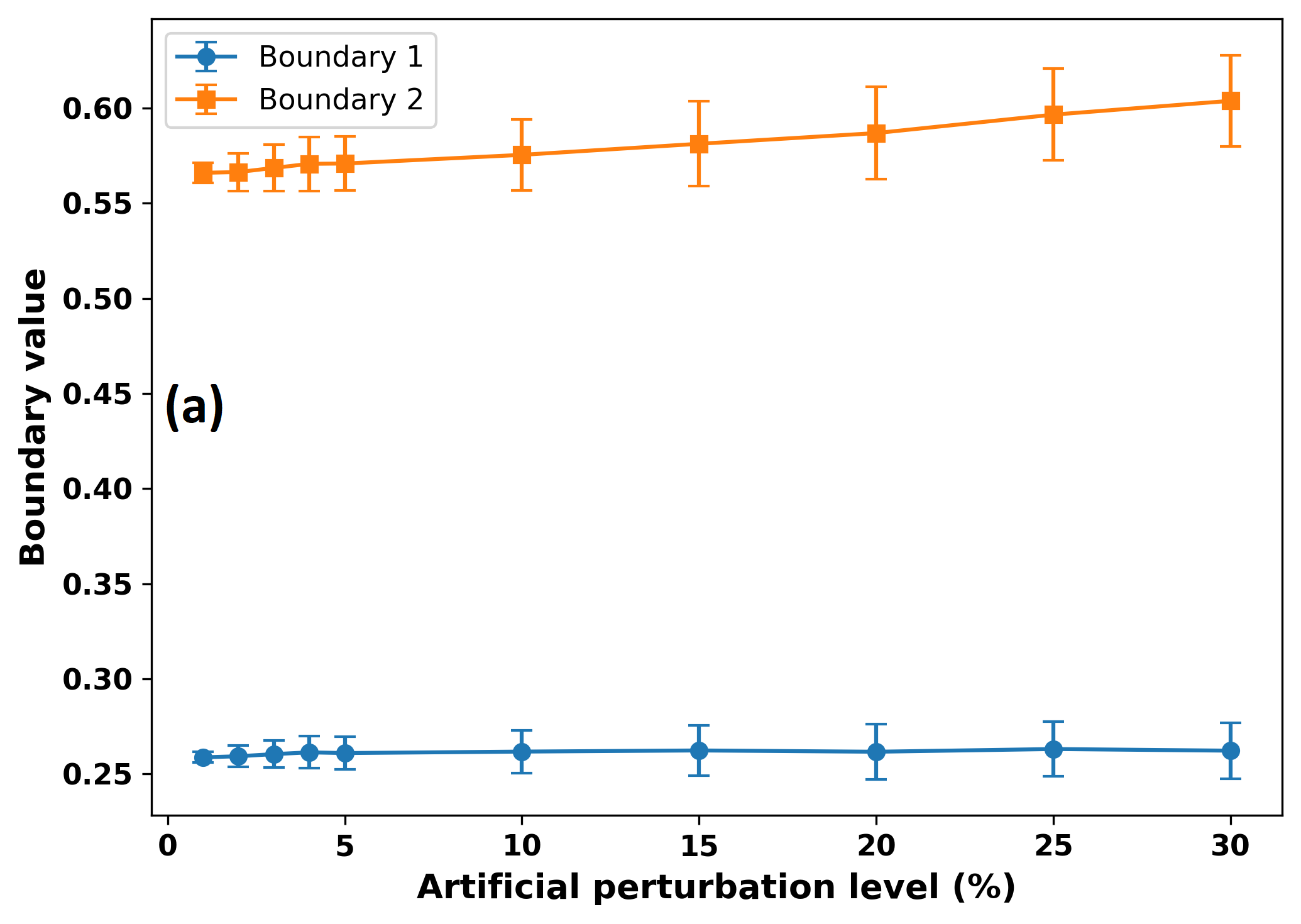}
\includegraphics[width=0.495\textwidth]{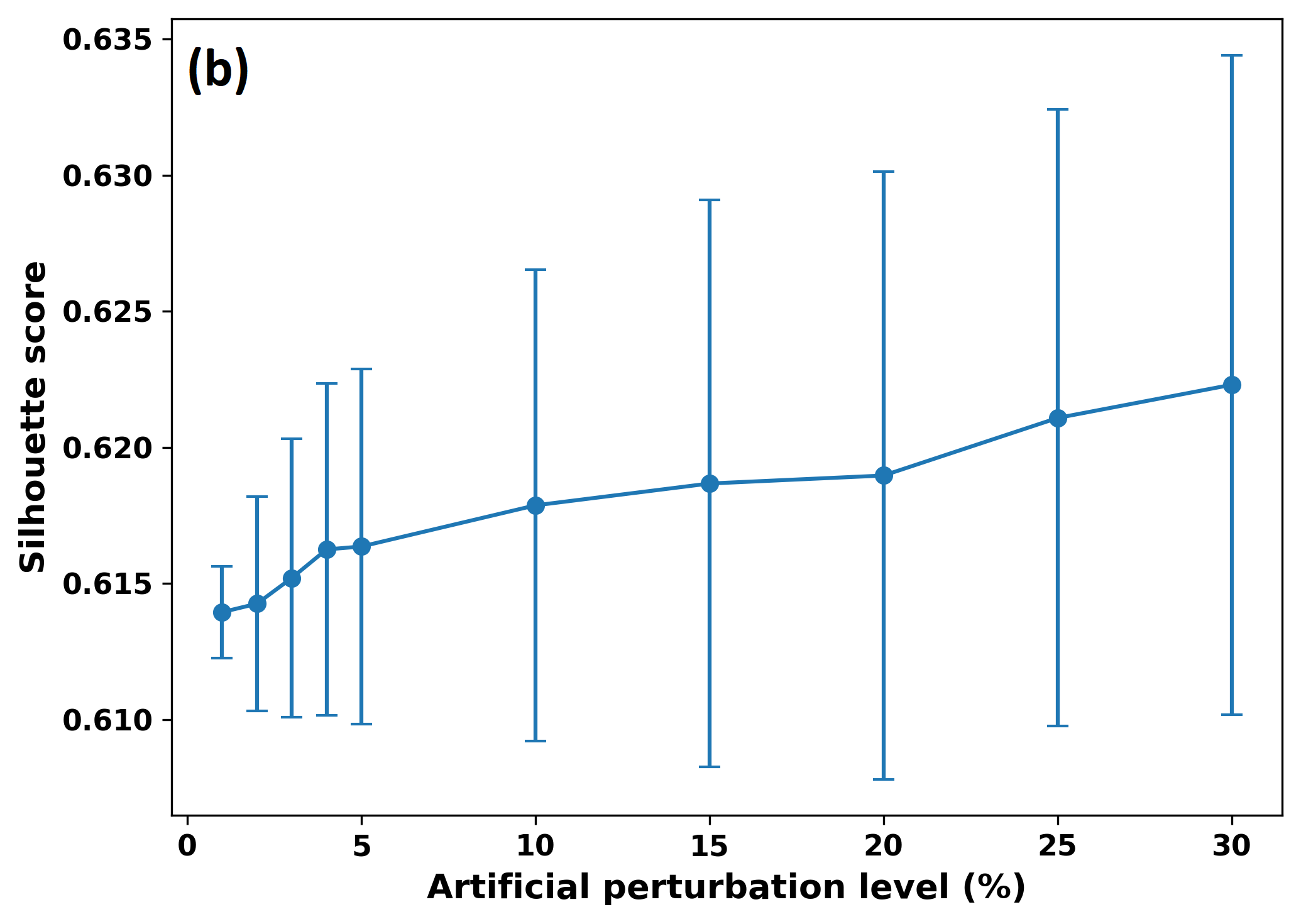}
\caption{(a) Evolution of the statistically derived fillout factor boundaries under progressively increasing artificial perturbations applied to the fillout factor distribution. Error bars represent the standard deviation of the derived boundaries over 500 realizations for each perturbation level. (b) Corresponding silhouette scores as a function of perturbation amplitude. The relatively stable silhouette values indicate that the global clustering structure remains preserved under moderate perturbations of the fillout factor distribution.}
\label{Fig:sensitivity}
\end{figure*}

Figure \ref{Fig:sensitivity} summarizes the results of all perturbation experiments. The lower boundary separating the shallow- and medium-contact subclasses remains remarkably stable across the entire perturbation range, varying only from $f \simeq 0.259$ to $f \simeq 0.263$. In contrast, the upper boundary separating the medium- and deep-contact subclasses exhibits a more gradual drift toward larger $f$ values as the perturbation amplitude increases, although the overall three-class structure remains preserved throughout the experiment. The standard deviations of both boundaries increase progressively with perturbation level, as expected for increasingly distorted realizations of the original distribution, but remain comparatively modest even at the largest perturbation amplitudes explored.

The silhouette scores also remain relatively stable over the full perturbation range, remained close to the original value of 0.613, varying only slightly from $\sim0.614$ to $\sim0.622$. This behavior indicates that the global clustering structure does not collapse under moderate perturbations of the fillout factor distribution. Figure \ref{Fig:sensitivity}a illustrates the evolution of the two statistical boundaries as a function of perturbation amplitude, while Figure \ref{Fig:sensitivity}b shows the corresponding silhouette scores and their associated dispersions across all realizations.

The derived subclass structure remains robust against moderate perturbations of the fillout factor distribution. The proposed subclass boundaries should therefore be interpreted primarily as statistically motivated divisions within a continuous evolutionary distribution rather than as strictly discrete physical states.

\subsection{Dependence of the statistical boundaries on sample composition}
The analyses presented in this study established a three-class statistical structure within the fillout factor distribution of contact binaries. While the sensitivity analysis demonstrated that these subclasses are stable under random perturbations of the $f$ distribution, an additional and necessary question is whether the inferred class boundaries depend on the physical composition of the underlying sample. We therefore examined whether the statistical segmentation of the fillout factor changes with sample composition. It is important to emphasize that this procedure does not aim to redefine the underlying distribution of the fillout factor, but rather to test the robustness of the clustering boundaries against changes in the composition of the underlying sample along independent physical parameters. This issue was investigated by dividing the full dataset into two subsets for each physical parameter ($P$, $q$, $T_1$, and $T_2$) using a median-based partition. The median split was adopted purely as a nonparametric and data-balanced criterion, ensuring comparable sample sizes in both subsets without imposing any physical threshold on the parameters. Each subset therefore represents systems located in the lower and upper segments of a given physical parameter. The clustering procedure described in the previous section was then independently applied to the fillout factor distribution of each subset.

The baseline classification yields two boundaries at $f = 0.257$ and $f = 0.561$, with a silhouette score of $0.613$. The corresponding results obtained from the subsampled populations are summarized in Table \ref{Tab:composition}. The differences between the baseline and subsampled cases are reported as $\Delta B_1$ and $\Delta B_2$, while $\Delta S$ denotes the corresponding change in silhouette score.

Table \ref{Tab:composition} indicates that the global clustering structure remains preserved across all parameter-conditioned subsets, as evidenced by consistently high silhouette values. However, shifts in the boundary positions are observed depending on the selected physical parameter. The most pronounced variations are found for the mass ratio, particularly affecting the second boundary separating the medium- and deep-contact systems, as summarized in Table \ref{Tab:composition}. This behavior is consistent with the strong association between the mass ratio and the fillout factor. In particular, the second boundary ($\Delta B_2$) exhibits the largest variation between the low- and high-value mass-ratio subsets. A similar behavior is observed for the orbital period and the effective temperatures. For these parameters, the corresponding boundary shifts are reported in Table \ref{Tab:composition}, showing moderate but systematic variations between the low- and high-value subsets. In all cases, the changes in silhouette score remain relatively small, typically within $\sim 0.04$ compared to the baseline value.

The proposed classification remains structurally stable in terms of cluster separability, while the precise location of the boundaries exhibits a moderate but parameter-dependent variation with the physical composition of the sample. The shallow-contact boundary is more stable across all tested conditions, whereas the second boundary shows a higher sensitivity to changes in the underlying parameter distribution. The proposed subclasses therefore form statistically meaningful partitions of the fillout factor distribution, although the exact boundary positions vary slightly with the physical properties of the sample and should not be regarded as universally fixed values.

\begin{table*}
\centering
\caption{Mean deviations of the class boundaries and silhouette score relative to the full-sample classification for different parameter-conditioned subsets.}
\begin{tabular}{lccc}
\hline
Parameter & $\Delta B_1$ & $\Delta B_2$ & $\Delta S$ \\
\hline
$P$(d)   & $-0.067 / -0.005$ & $-0.091 / -0.001$ & $-0.040 / -0.031$ \\
$q$   & $+0.048 / -0.087$ & $+0.049 / -0.134$ & $-0.040 / -0.038$ \\
$T_1$(K) & $-0.036 / -0.004$ & $-0.051 / 0.000$ & $-0.017 / -0.026$ \\
$T_2$(K) & $-0.064 / +0.028$ & $-0.081 / +0.028$ & $-0.036 / -0.039$ \\
\hline
\end{tabular}
\tablefoot{Each subset is defined using a median split of the corresponding physical parameter. $\Delta B_1$ and $\Delta B_2$ indicate the mean shifts of the class boundaries relative to the full-sample baseline values, and $\Delta S$ denotes the corresponding change in silhouette score, defined as $\Delta S = S_{\mathrm{subset}} - S_{\mathrm{full}}$. For each parameter, both low- and high-value subsets are reported.}
\label{Tab:composition}
\end{table*}

\subsection{Physical predictability of the fillout factor}
Following the clustering-based definition of the fillout factor classes, the present regression analysis examines whether the fillout factor can be predicted from the physical properties of contact binary systems, thereby providing a measure of its physical relevance. A RF regressor was trained using the available physical parameters ($P$, $q$, $r_1$, $r_2$, $T_1$, $T_2$) to map the multidimensional parameter space onto the fillout factor distribution. The resulting model achieves a coefficient of determination of $R^2 = 0.607$ and a mean absolute error of $0.088$, indicating that approximately $60.7\%$ of the variance in $f$ can be explained by the considered physical parameters. The dominant contribution to the regression model arises from the primary fractional radius, $r_1$, with an importance of 0.544, followed by the secondary radius $r_2$ with an importance of 0.170. This behavior is expected because the fractional radii and the fillout factor are both related to the Roche configuration of the binary system. Among the parameters that are not directly tied to the geometrical definition of $f$, the mass ratio provides the strongest predictive contribution, with an importance of 0.100. The thermal parameters contribute at a lower level, with $T_2$, $P$, and $T_1$ having importances of 0.079, 0.064, and 0.043, respectively.

The interpretation of the regression results requires some caution. The fractional radii are not independent of the fillout factor because, within the framework of Roche geometry, they are determined by the fillout factor and the mass ratio. More generally, the physical parameters of contact binaries are interconnected through mass transfer, AML, thermal coupling, and geometrical constraints (\citealt{1982A&A...109...17V}, \citealt{1995MNRAS.274.1019S}, \citealt{2008MNRAS.390.1577G}). Consequently, the reported feature importances should be interpreted as measures of predictive contribution rather than indicators of strictly independent physical effects.

To evaluate the extent to which the predictive performance is driven by geometrical quantities, the RF regression was repeated after excluding both fractional radii from the predictor set. Using only the orbital period, mass ratio, and component temperatures, the model achieves $R^2 = 0.331$ with a mean absolute error of 0.132. The substantial decrease in predictive accuracy relative to the full model demonstrates that a significant fraction of the predictive power originates from the geometrical information encoded in the fractional radii. Nevertheless, the model retains a non-negligible ability to predict the fillout factor, indicating that part of the observed relationship is also associated with physical parameters that are not directly linked to the Roche geometry. Within the reduced parameter set, the mass ratio becomes the dominant predictor, accounting for more than half of the total feature importance (0.510). The orbital period and secondary temperature contribute at comparable levels, with importances of 0.182 and 0.175, respectively, while the primary temperature contributes 0.132. The prominence of the mass ratio is consistent with its systematic variation across the fillout factor sequence and supports its central role in the long-term evolution of contact binary systems.

\begin{figure*}
\centering
\resizebox{\hsize}{!}{\includegraphics{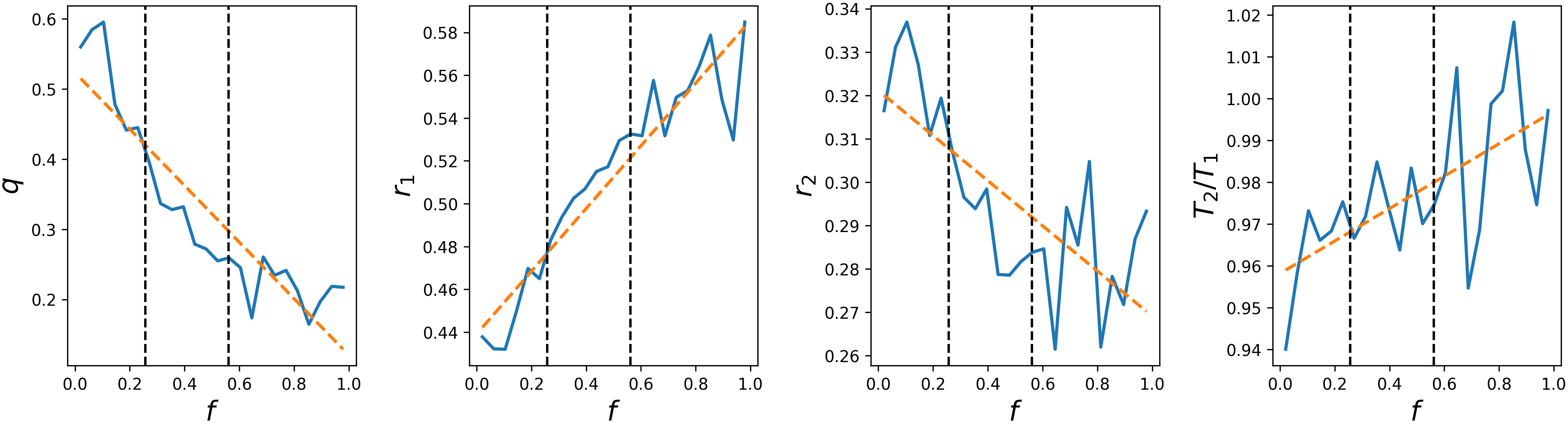}}
\caption{Mean values of the mass ratio, fractional radii, and temperature ratio as a function of the fillout factor. The curves represent binned averages, while the dashed vertical lines indicate the statistically derived boundaries between the shallow-, medium-, and deep-contact classes. The overall monotonic trends are quantified by the linear regression slopes discussed in the text.}
\label{Fig:fevol}
\end{figure*}

Further analysis was carried out to determine whether the fillout factor traces a coherent evolutionary sequence by quantifying monotonic trends of key physical parameters as a function of $f$ using linear regression slopes. This analysis complements the RF results by directly measuring the direction and magnitude of parameter changes across the fillout factor sequence. The mass ratio decreases systematically with increasing fillout factor ($\mathrm{slope} = -0.403$), while the primary fractional radius increases ($\mathrm{slope} = +0.146$). In contrast, the secondary fractional radius shows a weaker decreasing trend ($\mathrm{slope} = -0.052$), and the temperature ratio increases slightly ($\mathrm{slope} = +0.039$). Together, these trends are consistent with an evolutionary scenario involving mass redistribution, geometrical restructuring, and progressive thermal homogenization. Figure \ref{Fig:fevol} summarizes these evolutionary trends as a function of fillout factor, where the fitted slopes quantify the monotonic behavior of the principal physical parameters across the sample.

Although a substantial fraction of the variance in the fillout factor remains unexplained, the regression results demonstrate that $f$ is not merely a geometrical descriptor. Instead, it exhibits statistically significant connections with the physical properties of contact binaries, particularly the mass ratio, while the evolutionary trends reveal a coherent progression of system parameters across the fillout factor sequence.

\subsection{Thermal coupling across the fillout factor sequence}
The RF regression analysis presented in Section~3.3 showed that temperature-related parameters contribute less strongly to the prediction of the fillout factor than geometrical quantities such as the fractional radii and the mass ratio. Likewise, the Spearman correlation analysis presented in Section~2 revealed weaker monotonic associations between $f$ and the temperature parameters than between $f$ and the geometrical parameters. However, the comparatively low predictive importance of the thermal parameters does not necessarily imply that thermal evolution is unimportant in contact binaries. Thermal coupling remains a fundamental component of the common-envelope configuration and may still produce systematic trends along the fillout factor sequence. We therefore investigate whether signatures of progressive thermal homogenization are present across the statistically defined subclasses identified in Section~2.

To examine this possibility, we computed two quantities that directly characterize the thermal state of the binary components: the temperature difference, $|T_1-T_2|$, and the temperature ratio, $T_2/T_1$. These quantities were evaluated for the shallow-, medium-, and deep-contact subclasses defined by the statistically derived boundaries at $f=0.257$ and $f=0.561$. Unlike the parameter comparisons discussed in Section~2, which focused on the original physical parameters used in the classification analysis, the present investigation specifically targets the degree of thermal similarity between the two stellar components.

The mean temperature difference decreases systematically from approximately 319 K in the shallow-contact systems to 248 K in the medium-contact systems and 215 K in the deep-contact systems. At the same time, the mean temperature ratio increases from approximately $0.964$ to $0.974$ and $0.988$ across the same sequence. To quantify the statistical significance of these trends, additional Kruskal--Wallis tests were performed for both thermal indicators across the three fillout factor subclasses. The results confirm statistically significant differences for both $|T_1-T_2|$ ($H = 14.3$, $p = 7.7 \times 10^{-4}$) and $T_2/T_1$ ($H = 25.4$, $p = 3.1 \times 10^{-6}$).

The observed behavior suggests that systems with larger fillout factors tend to exhibit a greater degree of thermal homogenization between their stellar components. Nevertheless, the thermal trends remain noticeably weaker than the structural trends discussed in Section~3.3. For example, the monotonic evolution of the mass ratio and fractional radii across the fillout factor sequence produces substantially stronger statistical signatures than those associated with the temperature indicators. This difference is consistent with the RF results, where geometrical parameters dominate the predictive power of the model while temperature-related quantities provide only secondary contributions.

The substantial overlap of $|T_1-T_2|$ and $T_2/T_1$ between the subclasses further indicates that thermal equilibration remains incomplete even in many deep-contact systems. The thermal state of the common envelope therefore appears to evolve more gradually than the underlying geometrical configuration. Within the context of the present sample, the fillout factor sequence is primarily associated with structural evolution of the Roche geometry and mass distribution, while thermal homogenization emerges as a secondary trend superimposed on this broader evolutionary progression.

\subsection{Physical interpretation of the fillout factor sequence}
The identified trends suggest that variations in the fillout factor reflect systematic changes in the Roche geometry of contact binaries, consistent with the progressive evolution of the common-envelope configuration. The progressive decrease in the mass ratio toward larger fillout factors indicates that the transition from shallow to deep contact is accompanied by substantial changes in the binary configuration. Such behavior is consistent with evolutionary models of W UMa-type binaries undergoing mass transfer and AML.
This trend in the mass ratio evolution further supports the evolutionary interpretation of the fillout factor sequence discussed in this subsection. The mean mass ratio decreases from $q \simeq 0.52$ for shallow-contact systems to $q \simeq 0.22$ for deep-contact systems, suggesting progressive mass redistribution during long-term binary evolution. This interpretation is also in qualitative agreement with theoretical models of W UMa-type binary evolution. In particular, several evolutionary frameworks and TRO models (e.g., \citealt{2012JASS...29..145E}, \citealt{2012AcA....62..153S}, \citealt{2022ApJS..262...12K}, \citealt{2025ApJ...995...19F}) predict systematic relationships among the mass ratio, orbital period, and degree of contact. Accordingly, the statistically significant correlation between orbital period and fillout factor found in our sample is compatible with the observed $q$-$f$ relation and broadly consistent with these theoretical expectations. The comparatively strong dependence of the second statistical boundary on $q$ therefore suggests that the transition toward deep contact may be more strongly associated with variations in the mass ratio distribution of the analyzed sample. This interpretation is consistent with previous multivariate analyses indicating that the mass ratio acts as one of the principal evolutionary parameters governing the global structure and evolutionary state of contact binaries (\citealt{poro2026multivariate}). The classification analysis presented in Section 2 provides additional support for this interpretation. Also, as shown in Section 3.3, the mass ratio emerges as the most important predictor of the fillout factor classes among the nongeometrical parameters. The orbital period and component temperatures contribute at lower but comparable levels. Because the classification itself is defined exclusively from the fillout factor distribution, the prominence of the mass ratio cannot be attributed to a direct geometrical connection with $f$. Instead, it indicates that the fillout factor sequence is closely linked to systematic variations in the mass-ratio distribution. This result reinforces the interpretation that the observed subclasses trace an underlying evolutionary progression of contact binaries rather than merely representing a geometric description of the common-envelope configuration.

One of the most significant results of the present classification analysis is the asymmetric behavior of the two statistical boundaries. The transition between the shallow- and medium-contact subclasses remains remarkably stable under both perturbation experiments and parameter-conditioned subsampling, whereas the transition between the medium- and deep-contact subclasses exhibits noticeably larger variations. The differing stability of the two boundaries may reflect different physical trends along the contact sequence. In shallow-contact systems, relatively small changes in the Roche configuration may produce comparatively large structural responses because the common envelope is only weakly established. As a result, systems near the onset of contact occupy a more sharply defined region of parameter space. Deep-contact systems already possess a well-developed common envelope, allowing structural adjustments to occur more continuously over a broader range of fillout factors.

The comparatively gradual evolution observed in the deep-contact systems may also be connected to increasing thermal coupling between the stellar components. TRO models of contact binaries suggest that systems may evolve through alternating phases of stronger and weaker thermal disequilibrium within the common envelope (\citealt{1994ApJ...434..277W}, \citealt{2001MNRAS.328..914Q}, \citealt{stepien2006evolutionary}). The reduced statistical separation between the medium- and deep-contact subclasses in several physical parameters suggests that these parameters become less effective at distinguishing systems at larger fillout factors. Although the present analysis does not directly model energy transport processes, the observed statistical behavior appears broadly compatible with such an evolutionary picture. These results suggest that the absence of perfectly sharp statistical boundaries indicates that the transition between subclasses is likely continuous in physical terms, even though distinct clustering structures remain detectable within the observational parameter distribution.

\section{Conclusions}
This study presents a data-driven statistical classification of contact binary systems based on the fillout factor, using a sample of 796 contact stars. The results show that the distribution of $f$ naturally separates into three statistically distinguishable subclasses corresponding to shallow-, medium-, and deep-contact configurations, with boundaries located at $f \simeq 0.257$ and $f \simeq 0.561$. The clustering solution yields a stable silhouette score of $S \simeq 0.613$, indicating that the fillout factor distribution contains a statistically meaningful, although continuous, internal structure.

The robustness of this structure was examined through both artificial perturbation experiments and parameter-conditioned subsampling. Under artificial perturbations of up to 30\% in the fillout factor distribution, the lower boundary remains comparatively stable while the upper boundary shows a more noticeable drift. Despite these variations, the overall clustering structure is preserved, supporting the stability of the identified subclasses. Although the three-class structure is retained across all parameter-conditioned populations, the second boundary exhibits systematically larger variations than the first. This asymmetry suggests that the transition toward deeper contact may be more sensitive to the physical composition of the underlying sample than the onset of contact itself. It is also noteworthy that the statistically derived boundaries lie close to the empirical thresholds commonly adopted in the literature, despite being obtained without imposing any predefined classification limits (\citealt{2022AJ....164..202L}, \citealt{2023MNRAS.519.5760L}, \citealt{2020RAA....20..163Q}).

The regression analysis further indicates that the fillout factor is not an entirely independent quantity, but is statistically linked to several observable properties of contact binaries. The RF regression achieves $R^2 = 0.607$ with a mean absolute error of 0.088, indicating that a substantial fraction of the observed variance in $f$ is associated with measurable system properties. The RF classification analysis indicates that the mass ratio is the strongest predictor of the statistically derived fillout factor subclasses, followed by the orbital period and component temperatures. This result strengthens the interpretation that the fillout factor sequence is linked not only to the geometric state of the common envelope but also to the underlying physical evolution of the binary system.

The observed decrease in mass ratio with increasing fillout factor is qualitatively consistent with evolutionary scenarios in which mass transfer and AML gradually drive contact binaries toward deeper contact configurations (\citealt{1982A&A...109...17V}, \citealt{1995MNRAS.274.1019S}, \citealt{2005ApJ...629.1055Y}, \citealt{2025ApJ...995...19F}). Combined with the systematic variation in the physical parameters across the fillout factor sequence and the increased sensitivity of the medium-to-deep boundary to sample composition, these findings are consistent with an evolutionary interpretation of the derived subclasses. The subclasses should therefore be interpreted as statistically distinguishable segments of a continuous distribution in the fillout factor, rather than discrete physical states.

The present analysis should be considered in the context of several limitations associated with the available observational and modeling data. The reported fillout factors were compiled from studies employing different light curve modeling codes and heterogeneous modeling strategies, which may introduce systematic differences in the derived values of $f$ from one system to another. Furthermore, the quality and precision of the underlying photometric observations are not uniform across the sample and may contribute additional uncertainties to the derived parameters. The lack of uniformly available spectroscopic constraints also limits the ability to fully disentangle geometric and physical effects in some systems. Although these limitations cannot be fully mitigated within the scope of the present work, they do not alter the main statistical conclusions.

\begin{acknowledgements}
We would like to express our sincere appreciation to the referee for the valuable comments and suggestions, which have helped us improve the quality of the manuscript.
The authors also acknowledge the publicly available datasets that enabled this work. The analysis was carried out using Python-based scientific computing libraries for statistical analysis, machine learning, and visualization. We sincerely appreciate the Raderon AI Laboratory (\url{https://raderonlab.ca}) and Soroush Sarabi for providing computational facilities used in this study.
\end{acknowledgements}

\begin{appendix}
\onecolumn
\section{Sample distribution}

\begin{figure*}[ht!]
\resizebox{\hsize}{!}{\includegraphics{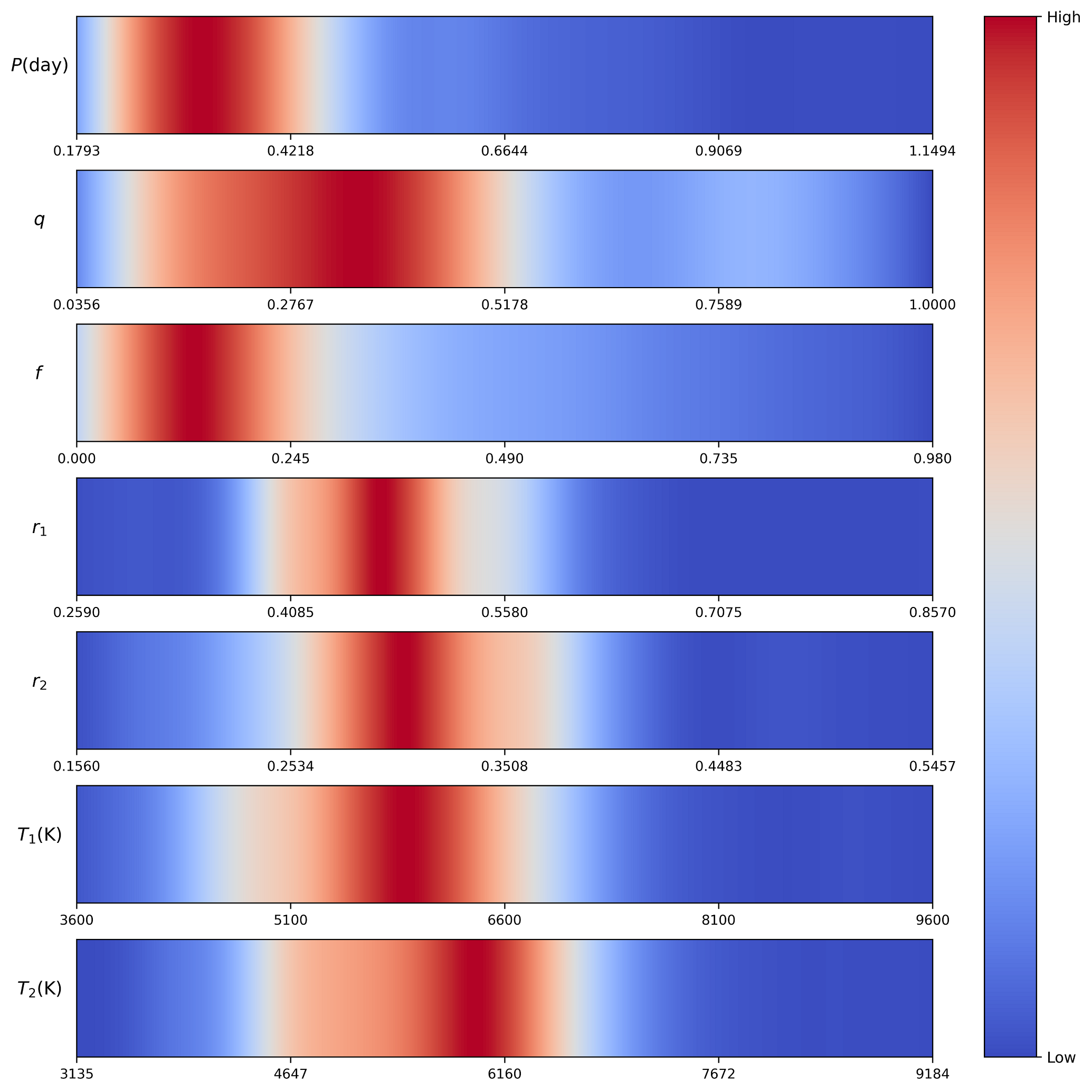}}
\caption{Density-strip representation of the distributions of the principal physical and geometrical parameters for the 796 contact binary systems used in this study. The colors represent kernel density estimates, with warmer colors indicating regions of higher data density.}
\label{Fig:density}
\end{figure*}

\begin{figure*}[ht!]
\resizebox{\hsize}{!}{\includegraphics{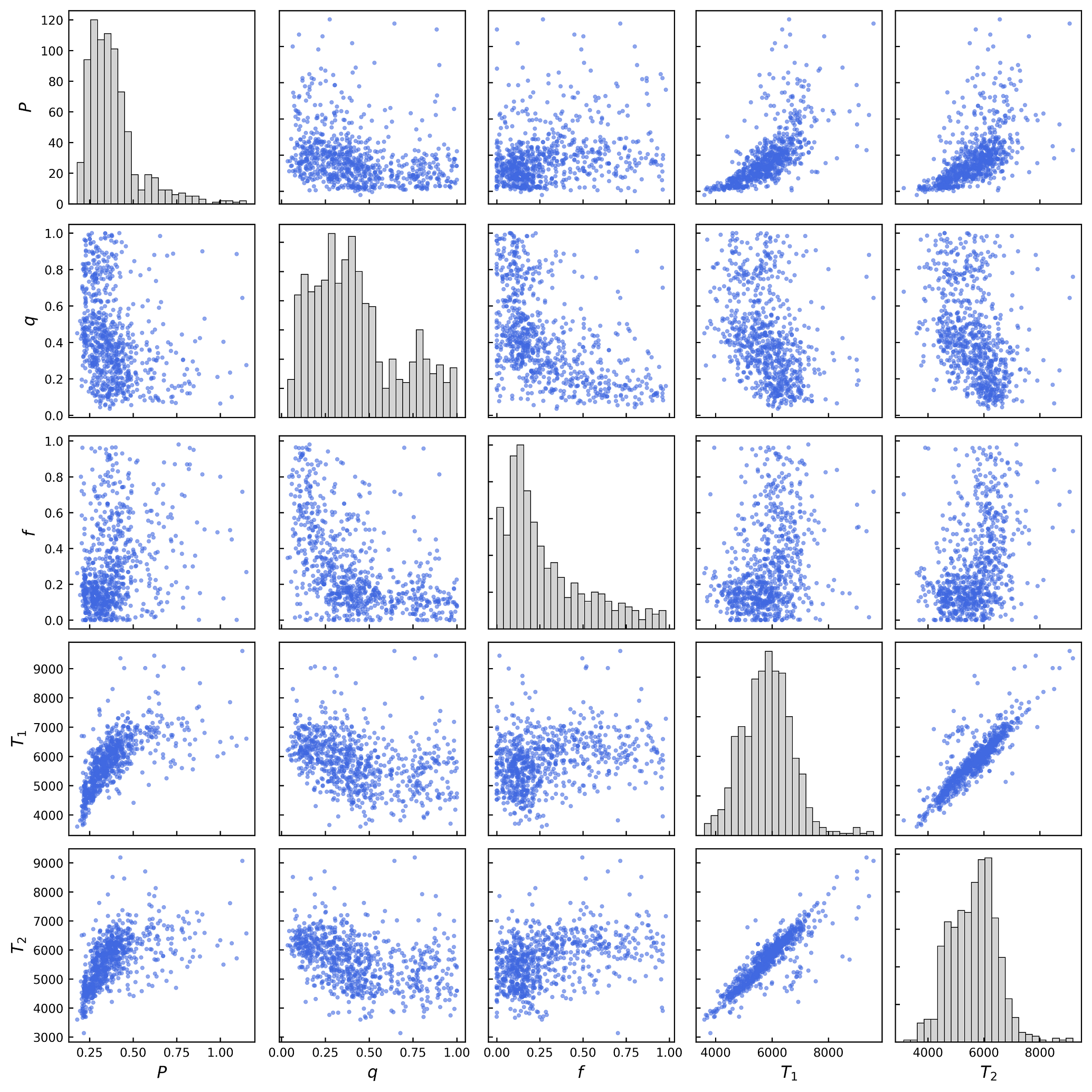}}
\caption{Pairwise scatter matrix of the principal parameters for the 796 contact binary systems included in the statistical sample.}
\label{Fig:scatter}
\end{figure*}

\end{appendix}


\begin{thebibliography}{}
\bibitem[Breiman(2001)]{2001MachL..45....5B} Breiman L., 2001, MachL, 45, 5
\bibitem[Cui et al.(2012)]{2012RAA....12.1197C} Cui X.-Q., Zhao Y.-H., Chu Y.-Q., Li G.-P., Li Q., et al., 2012, RAA, 12, 1197
\bibitem[Eggleton(2012)]{2012JASS...29..145E} Eggleton P.~P., 2012, JASS, 29, 145
\bibitem[Fabry \& Pr{\v{s}}a(2025)]{2025ApJ...995...19F} Fabry M., Pr{\v{s}}a A., 2025, ApJ, 995, 19
\bibitem[Flannery(1976)]{1976ApJ...205..217F} Flannery B.~P., 1976, ApJ, 205, 217
\bibitem[Gazeas \& St{\c{e}}pie{\'n}(2008)]{2008MNRAS.390.1577G} Gazeas K., St{\c{e}}pie{\'n} K., 2008, MNRAS, 390, 1577
\bibitem[\protect\citeauthoryear{Kobulnicky et al.}{2022}]{2022ApJS..262...12K} Kobulnicky H.A., Molnar L.A., Cook E.M., Henderson L.E., 2022, ApJS, 262, 12
\bibitem[Kuiper(1941)]{1941ApJ....93..133K} Kuiper G.~P., 1941, ApJ, 93, 133
\bibitem[Li et al.(2022)]{2022AJ....164..202L} Li K., Gao X., Liu X.-Y., Gao X., Li L.-Z., et al., 2022, AJ, 164, 202
\bibitem[Li, Han, \& Zhang(2004)]{2004MNRAS.355.1383L} Li L., Han Z., Zhang F., 2004, MNRAS, 355, 1383
\bibitem[Liu et al.(2023)]{2023MNRAS.519.5760L} Liu X.-Y., Li K., Michel R., Gao X., Gao X., Liu F., Yin S.-P., et al., 2023, MNRAS, 519, 5760
\bibitem[Liu, Qian, \& Xiong(2018)]{2018MNRAS.474.5199L} Liu L., Qian S.-B., Xiong X., 2018, MNRAS, 474, 5199
\bibitem[Lucy(1968)]{1968ApJ...151.1123L} Lucy L.~B., 1968, ApJ, 151, 1123
\bibitem[Lucy(1976)]{1976ApJ...205..208L} Lucy L.~B., 1976, ApJ, 205, 208
\bibitem[Luo et al.(2015)]{2015RAA....15.1095L} Luo A.L., Zhao Y.H., Zhao G., Deng L.-C., Liu X.-W., et al., 2015, RAA, 15, 1095
\bibitem[Paki, Poro, \& Moosavi Rowzati(2025)]{2025Galax..13...74P} Paki E., Poro A., Moosavi Rowzati M.~D., 2025, Galax, 13, 74
\bibitem[Poro et al.(2025)]{2025MNRAS.538.1427P} Poro A., Jahangiri E., Sarvari E., Aliakbari R., et al., 2025, MNRAS, 538, 1427
\bibitem[Poro et al.(2026)]{poro2026multivariate} Poro A., Poggiani R., Foroutanfar A., Harzandjadidi R., Kahali Poor N., Alicavus, F., 2026, A\&A, 710, A49
\bibitem[Pr{\v{s}}a \& Zwitter(2005)]{2005ApJ...628..426P} Pr{\v{s}}a A., Zwitter T., 2005, ApJ, 628, 426
\bibitem[Pr{\v{s}}a et al.(2016)]{2016ApJS..227...29P} Pr{\v{s}}a A., Conroy K.E., Horvat M., Pablo H., et al., 2016, ApJS, 227, 29
\bibitem[Qian \& Yang(2004)]{2004AJ....128.2430Q} Qian S.-B., Yang Y.-G., 2004, AJ, 128, 2430
\bibitem[Qian et al.(2020)]{2020RAA....20..163Q} Qian S.-B., Zhu L.-Y., Liu L., Zhang X.-D., et al., 2020, RAA, 20, 163
\bibitem[Qian(2001)]{2001MNRAS.328..914Q} Qian S., 2001, MNRAS, 328, 914
\bibitem[St{\k{e}}pie{\'n}(1995)]{1995MNRAS.274.1019S} St{\k{e}}pie{\'n} K., 1995, MNRAS, 274, 1019
\bibitem[St{\k{e}}pie{\'n}(2006)]{stepien2006evolutionary} St{\k{e}}pie{\'n} K., 2006, Acta Astronomica, 56, 199
\bibitem[\protect\citeauthoryear{St{\k{e}}pie{\'n} \& Gazeas}{2012}]{2012AcA....62..153S} St{\k{e}}pie{\'n} K., Gazeas K., 2012, AcA, 62, 153
\bibitem[Vilhu(1982)]{1982A&A...109...17V} Vilhu O., 1982, A\&A, 109, 17
\bibitem[Wang(1994)]{1994ApJ...434..277W} Wang J.-M., 1994, ApJ, 434, 277
\bibitem[Wilson \& Devinney(1971)]{1971ApJ...166..605W} Wilson R.~E., Devinney E.~J., 1971, ApJ, 166, 605
\bibitem[Wilson(1979)]{1979ApJ...234.1054W} Wilson R.~E., 1979, ApJ, 234, 1054
\bibitem[Yakut \& Eggleton(2005)]{2005ApJ...629.1055Y} Yakut K., Eggleton P.~P., 2005, ApJ, 629, 1055
\end{thebibliography}
\end{document}